# TIME-DRIFT AWARE RF OPTIMIZATION WITH MACHINE LEARNING TECHNIQUES*

R. Sharankova†, M. Mwaniki, K. Seiya, M. Wesley
Fermi National Accelerator Laboratory, Batavia, IL 60510, USA

*Abstract*

The Fermilab Linac delivers 400 MeV H- beam to the rest of the accelerator chain. Providing stable intensity, energy, and emittance is key since it directly affects downstream machines. To operate high current beam, accelerators must minimize uncontrolled particle loss; this can be accomplished by minimizing beam longitudinal emittance via RF parameter optimization. However, RF tuning is required daily since the resonance frequency of the accelerating cavities is affected by ambient temperature and humidity variations and thus drifts with time. In addition, the energy and phase space distribution of particles emerging from the ion source are subject to fluctuations. Such drift is not unique to Fermilab, but rather affects most laboratories. We are exploring machine learning (ML) algorithms for automated RF tuning for 2 objectives: optimization of Linac output energy and phase oscillation correction, with an emphasis on time-drift aware modeling that can account for conditions changing over time.

## THE FERMILAB LINAC

The Fermi National Accelerator Laboratory (Fermilab) Linac accelerates H- beam to 401.5 MeV. The Linac is preceded by a 35 keV H- ion source and a pre-accelerator which bunches and accelerates beam to 750 keV. The Linac comprises three sections: a Drift Tube Linac (DTL), a Side Coupled Linac (SCL), and a transition section between them. The DTL comprises 207 drift tubes spread across 5 tanks and operates at 201.25 MHz RF frequency. The SCL has 7 modules with 448 total cells, operating at 805 MHz. The transition section consists of a buncher and a vernier cavity for longitudinal matching between the DTL and the SCL. During regular operations, the Linac delivers roughly 25 mA at 35 $\mu$s pulse length with transition efficiency $\geq$ 92%.

## LINAC RF & LLRF

DTL RF field amplitude is controlled by the Marx modulator logic controller which in turn controls the 5 MW power tube modulator voltage [1]. The RF phase is controlled by the low level RF (LLRF) module in a VME eXtension for Instrumentation (VXI) crate. Each SCL module is powered by a 12 MW klystron with VXI based LLRF phase and amplitude control. Amplitude and phase settings are send to the front-end card in the LLRF VXI crates via the Fermilab Accelerator control network (ACNET) [2].

* This material is based upon work supported by the U.S. Department of Energy, Office of Science, Office of High Energy Physics, under contract number DE-AC02-07CH11359.
† rshara01@fnal.gov

## LINAC DAILY TUNING

Stable Linac output is crucial for downstream machines. Drifting resonance frequencies of the accelerating cavities and fluctuations in the beam energy and phase coming from the ion source and pre-accelerator both induce longitudinal emittance growth and increased particle loss. To counter such effects, operators perform machine tuning several times a day by hand-scanning a handful of RF parameters. The two main objectives of this tuning are (1) optimize Linac output energy, and (2) maximize beam throughput while minimizing beam losses along Linac.

## RF OPTIMIZATION WITH ML

Hand-tuning faces several challenges. Human operators cannot optimize in multi-dimensional space, but rather scan parameters one by one making it difficult to ensure the system has reached a global optimum. Tuning is limited by personnel availability, rather than done when Linac conditions change. To resolve these challenges, we are working towards automating the tuning procedure using ML-based algorithms. We have developed two independent approaches for the two tuning objectives outlined in the previous section.

### Linac output energy control

After leaving the Linac, beam is injected into a rapid cycling synchrotron called the Booster via a transfer line. Central momentum changes from the design output energy of 401.5 MeV or increased momentum spread (longitudinal emittance) can manifest as radial position errors in the Booster, increasing beam loss. Thus it is vital to deliver stable Linac output energy that matches Booster acceptance to minimize loss in downstream machines. In daily tuning

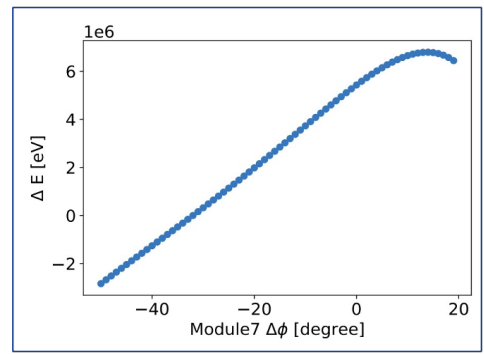

Figure 1: Change in Linac output energy as a function of SCL7 RF phase change (simulation).

Linac output energy is optimized by hand-scanning the RF cavity phase of the last Linac cavity, SCL7. Figure 1 shows in simulation how output energy changes as a function of cavity phase change.

Fluctuations in the output energy can be measured using beam position monitor (BPM) data in the transfer line from Linac to Booster where there are no accelerating elements, only focusing and bending, and beam is drifting. Using the correlation between transverse displacement $\Delta x$(or $\Delta y$) and centroid momentum change $\Delta p/p$ in regions with dispersion, one can write $\Delta x(\Delta y) = D_{x(y)} \Delta p/p$ where $D_{x(y)}$ is a dispersion coefficient derived from MAD-X [3] lattice simulation [4]. Of particular interest to this study are 3 horizontal BPMs in locations along the injection line where dispersion coefficients are large. In order of distance from SCL7 these are called HPQ3, HPQ4 and HPQ5.

The change in cavity output energy (or transverse positions as a proxy) as a function of RF phase is defined here as RF cavity response. It was observed that response is not constant, but rather drifts with time, as shown in Figure 2 in terms of HPQ4 data. It is clear output energy depends on beam input energy at cavity entrance.

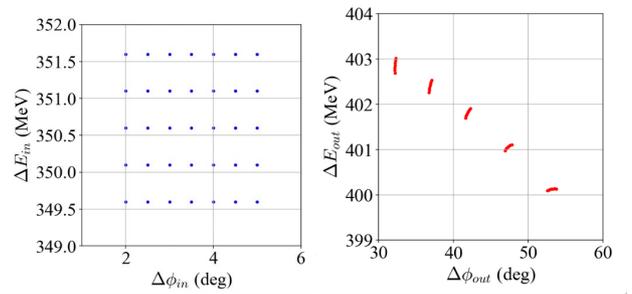

Figure 3: Simulated beam phase and energy relative to the baseline model at entrance (left) and exit (right) of SCL7.

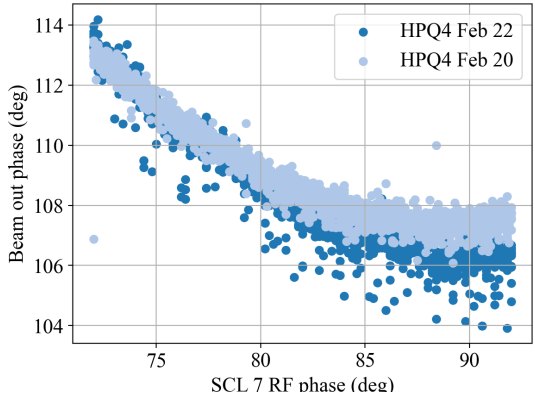

Figure 2: Beam horizontal position deviation at location HPQ4 as a function of SCL7 RF phase. Light blue trace is data from Feb 20, 2023. Dark blue trace is from Feb 22.

As a step towards automating Linac output energy control, we developed a ML-based procedure for 1-step energy drift correction that does not require daily SCL7 RF scanning. As a starting point for our modeling, we collected a reference SCL7 phase-scan dataset on February 22 2023 recording BPM transverse positions in the transfer line. We fitted a simulated RF cavity response curve to the reference data, by changing cavity voltage, input energy and input phase, to obtain baseline response. Next, the input phase-space dependence of the RF cavity response was modeled using simulation. Beam phase-space at the entrance of SCL7 is painted around the baseline as shown in Figure 3 (left) to cover all possible daily fluctuations, and phase-space at cavity exit is simulated (same Figure on the right).

A fully-connected deep neural network (DNN) was trained to predict beam input energy relative change from the baseline $\Delta E_{in}$ given input and output relative phase changes $\Delta\phi_{in}$, $\Delta\phi_{out}$, and output energy change $\Delta E_{out}$. The network was developed using the Tensorflow framework [5]. The network has 5 hidden Dense layers with ReLU activation and 20 nodes each. The optimizer used was Adam with initial learning rate of 0.001 and a custom rate decay scheduler. The network was trained on a sample of 1200 simulated examples for 20 epochs.

The trained model can be used on data to correct output energy daily drift by computing the relative differences w.r.t. the reference dataset. Input and output phases are measured with BPMs directly upstream and downstream of SCL7, and output energy is measured using transverse positions as outlined above. The model predicted $\Delta E_{in}$ for that particular day is then used to compute a calibrated RF cavity response curve in simulation. Finally, the calibrated RF cavity response is used to determine what SCL7 RF phase shift needs to be applied to the daily data to return the output energy to the reference 401.5 MeV. The correction mechanism was tested on data collected on Feb 20 2023. The effect of the RF phase correction on BPM horizontal positions at HPQ3, HPQ4 and HPQ5 is shown in Figure 4.

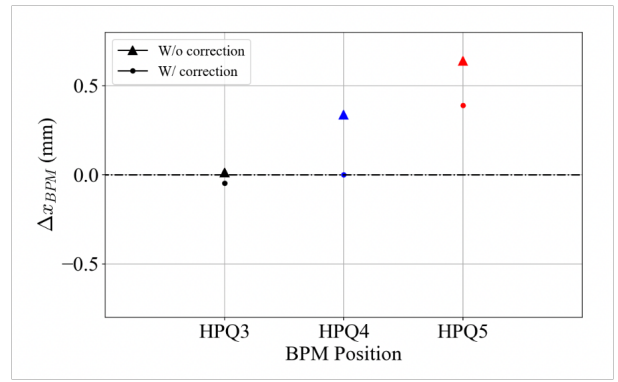

Figure 4: Effect of energy correction on beam horizontal positions at HPQ3, HPQ4 and HPQ5 locations.

It is clear the correction brings all horizontal positions closer to the reference. However one can see that the effect is slightly different at the 3 BPM locations. Investigating the raw BPM response vs. cavity phase for that day, we see that the shape of the 3 response curves is slightly different. This can be explained by possible errors on the calculated dispersion factors or horizontal displacements. The former are simulated assuming ideal beam trajectory and design bending magnet field strengths, and are thus susceptible to fluctuations in magnet currents and beam transverse motion. The latter are affected by the intrinsic noise on BPM positions of ≈ 0.1 mm [6]. Nevertheless, the correction resolution is better than the observed daily drift of ≥ 0.3 MeV.

*Phase oscillation correction*

Changes in beam energy or phase upstream of the Linac result in longitudinal emittance growth and will manifest as a beam phase oscillation propagating through the linac. The resulting elevated losses degrade beam throughput, thus correcting them is important for best Linac performance. Figure 5 shows evolution over 10 hours of beam phase as measured by BPMs without any changes to Linac RF parameters.

From simulation we expect ∼ 7 synchrotron oscillations in DTL, however there are only 5 BPMs in DTL, starting at DTL2 exit. Due to lack of sufficient instrumentation in the pre-accelerator and DTL, we cannot determine where the oscillations starts. However we have robust instrumentation coverage in the SCL, which allows us to measure the oscillation there well. Thus we developed a correction scheme that aims to correct the phase oscillation as seen by SCL BPMs.

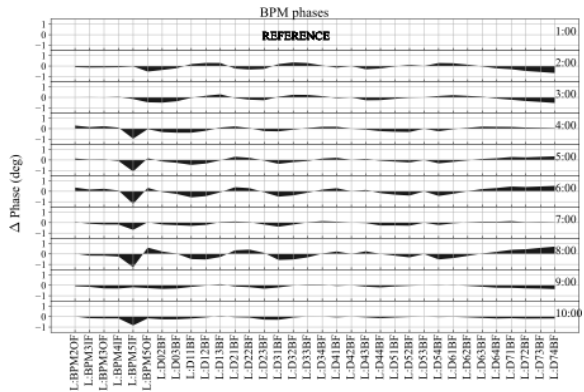

Figure 5: BPM phase evolution along Linac over 10 hours.

To counter the effect in SCL of a phase or energy error occurring upstream, we need to apply two independent (non-parallel) corrections in the DTL. Figure 6 (left) demonstrates in simulation the effect on beam phase of two independent RF shifts to DTL cavities. Same figure shows on the right the phase-space at the exit of the DTL (≈ 77 m in distance) painted by a 2D RF phase scan of the two cavities.

We developed a ML-based phase oscillation correction scheme based on the above principle. First, a 2D RF phase scan of DTL2 and DTL5 cavities was performed and BPM phase data was recorded. Then a fully-connected DNN was trained to reconstruct DTL2 and DTL5 RF phase from the phase oscillation pattern observed by 32 BPMs, 5 in the DTL and 27 in the SCL. The network has 10 hidden layers in total: 5 Dense layers with ReLU activation of 32,64,128,64 and 32 nodes, respectively, each followed by a Batch Normalization layer. Adam with initial learning rate of 0.001 and a custom rate decay scheduler was used for optimization. Network was trained on a sample of 910 data examples for 600 epochs.

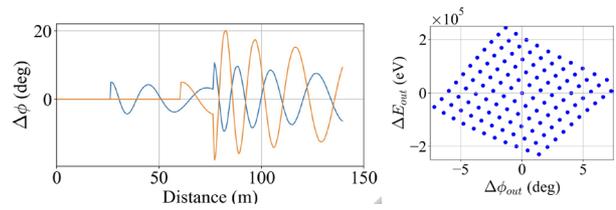

Figure 6: Effect of two independent RF shifts in DTL on beam phase (left). Beam phase-space at exit of DTL resulting from a 2D RF phase scan in DTL (right).

The model's ability to reconstruct arbitrary phase oscillation patterns by a combination of DTL and DTL5 RF phase shifts was tested on dataset from Dec 14 2022 where beam phase at Linac entrance is intentionally changed by different magnitude. Several examples of model performance are shown in Figure 7. It is clear that the model is qualitatively

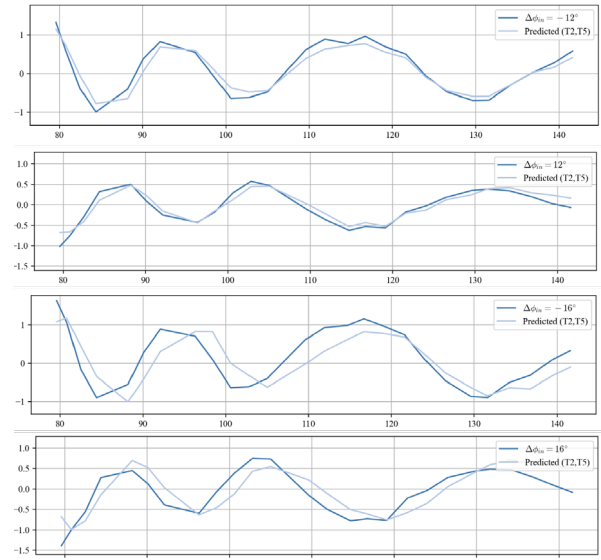

Figure 7: BPM phase oscillation along SCL. Horizontal axis is distance (m) Dark blue: effect of beam phase change upstream of Linac. Light blue: model predicted DTL2, DTL5 phase shifts.

able to reproduce a wide range of upstream oscillations, making oscillation correction possible.

## CONCLUSION

We are exploring ML applications using diagnostics data for automated Linac RF parameter optimization. We developed ML algorithms for two RF optimization objectives: Linac output energy control, and Linac phase oscillation correction. Preliminary results for both approaches are very promising. Future work includes testing of different neural network types to compare performance, and testing model prediction in real-time in the accelerator operations.